\documentclass[aps,pra,reprint, amsmath, amssymb,superscriptaddress,nofootinbib]{revtex4-1}
\usepackage{babel}
\usepackage{bm}
\usepackage[retainorgcmds]{IEEEtrantools}
\usepackage{graphicx}
\usepackage{mathrsfs}
\usepackage{amsmath}
\usepackage{amssymb}
\usepackage{color}
\usepackage{amsfonts}
\usepackage{times,txfonts}
\usepackage{nicefrac}
\usepackage[colorlinks=true,linkcolor=blue,urlcolor=blue,citecolor=blue,pdfusetitle]{hyperref}
\usepackage{amsmath}
\DeclareMathOperator{\sign}{sign}

\begin{document}

\title{Thermodynamic uncertainty relation from involutions}
\date{\today}
\author{Domingos S. P. Salazar}
\affiliation{Unidade de Educa\c c\~ao a Dist\^ancia e Tecnologia,
Universidade Federal Rural de Pernambuco,
52171-900 Recife, Pernambuco, Brazil}

\begin{abstract}

The Thermodynamic Uncertainty Relation (TUR) is a lower bound for the variance of a current as a function of the average entropy production and average current. Depending on the assumptions, one obtains different versions of the TUR. For instance, from the exchange fluctuation theorem, one obtains a corresponding exchange TUR. Alternatively, we show that TURs are a consequence of a very simple property: every process $s$ has only one inverse $s'=m(s)$, where $m$ is an involution, $m(m(s))=s$. This property allows the derivation of a general TUR without using any fluctuation theorem. As applications, we obtain the exchange TUR, the asymmetric TUR, a fluctuation-response inequality and lower bounds for the entropy production in terms of other nonequilibrium metrics.
\end{abstract}
\maketitle{}

%\section{Introduction}

% {\bf \emph{Introduction -}}
{\bf \emph{Introduction -}} 
In nonequilibrium thermodynamics, physical observables fluctuate in time, such as currents $\phi$ of particles or heat \cite{RevModPhys.93.035008,Seifert2012Review,Bustamante2005,Esposito2009,Jarzynskia2008,Jarzynski1997,Crooks1998,Gallavotti1995,Evans1993}. These fluctuations are fully encoded in the probability density function $p(\sigma,\phi)$, where $\sigma$ is the entropy production, which is usually system dependent and time dependent in transient regimes. However, there is something apparently universal in the fluctuations of any current $\phi$: the variance of $\phi$ (normalized by $\langle \phi \rangle^2$) is lower bounded by a function of the average entropy production. This is called the Thermodynamic Uncertainty Relation (TUR) \cite{Barato2015A,Gingrich2016,Polettini2017,Pietzonka2017,Hasegawa2019,Timpanaro2019b,Horowitz2022,Potts2019,Proesmans2019}, usually written as
\begin{equation}
\label{TUR}
    \frac{\langle \phi^2 \rangle - \langle \phi \rangle^2}{\langle \phi \rangle^2} \geq 
    f(\langle \sigma \rangle),
\end{equation}
where $f(x)$ is a function that depends on some characteristics of the system. The TUR was first derived for Markov jump processes \cite{Gingrich2016}, where $f(x)=2/x$. Alternatively, if the system satisfies the exchange fluctuation theorem (XFT), $p(\phi\,\sigma)=e^\sigma p(-\phi,-\sigma)$, then $f(x)$ has another particular form \cite{Timpanaro2019b,Hasegawa2019}.

In systems that are not time symmetric, XFT does not hold, but a different form of asymmetric TUR was obtained from the definition of entropy production, $p(\sigma,\phi)=e^\sigma \overline{p}(-\sigma,-\phi)$ \cite{Potts2019}, where $\overline{p}$ is the probability function in the backward experiment. In a similar setup, it was derived a Hysteretic TUR \cite{Proesmans2019}, with a tighter form proposed in \cite{Gianluca2022}.

In both cases, some form of fluctuation theorem (FT) was assumed explicitly. In this paper, we ask if there is anything more fundamental than the FT behind the derivation of the TURs. We found that TURs can be understood as a consequence of the fact that each process $s$ has a single inverse $s':=m(s)$, which gives $m(m(s))=s$. This map $m$, which is called an involution, can be explored to create the involution TUR (iTUR). In the most simple form, it reads
\begin{equation}
\frac{\langle \phi^2 \rangle - \langle \phi \rangle^2}{\langle \phi \rangle^2} \geq f(D(P|P')),
\end{equation}
valid for any asymmetric current, $\phi(s')=-\phi(s)$, and any involution $m$, where $P'(s):=P(m(s))$ and $D(P|P')=\sum_s P(s)\ln(P(s)/P'(s))$ is the Kullback-Leibler (KL) divergence, $f(x)=\sinh^{-2}(g(x)/2)$, and $g(x)$ the inverse of $h(x)=x\tanh(x/2)$, for $x\geq0$.

The paper is organized as follows. First, we present the formalism and our main result. Then, prove the main result based only on the involution property. After that, we discuss the result and apply it in the context of the exchange FT to obtain the exchange TUR. We also apply it to the framework of stochastic thermodynamics with asymmetric protocols, obtaining a tight version of the asymmetric TUR. Finally, we show how the result is useful to connect the KL divergence with the total variation distance and a general relation for the moment generating function. We also obtain a fluctuation-response inequality and the Crámer-Rao bound in a limiting case.

{\bf \emph{Formalism -}} We denote $s$ the elements of a set $S$. In the context of thermodynamics, each $s$ is a process that is observed with probability $P(s)\in [0,1]$. Additionally, we consider an involution $m:S\rightarrow S$, such that $s':=m(s)$ and
\begin{equation}
\label{involution}
m(m(s))=s.
\end{equation}
It means that each process has only one inverse, which is the simplest property one might demand from the arrow of time. We define $P':S\rightarrow [0,1]$ as 
\begin{equation}
\label{Pprime}
P'(s):=P(s')=P(m(s)),
\end{equation}
using a notation introduced in \cite{Salazar2022b}. Similarly, $Q:S\rightarrow [0,1]$ is another probability function and $Q'(s)=Q(s')=Q(m(s))$. Normalization reads $\sum_s P(s) = \sum_{s} P'(s) = \sum_s Q(s) = \sum_{s}Q'(s)=1$.
Let $\phi:S\rightarrow \mathbb{R}$ be any current with property
\begin{equation}
\label{current}
\phi(m(s))=-\phi(s),
\end{equation}
which makes it asymmetric under the involution. Consider the usual notation for the mean, $\langle \phi \rangle_P = \sum_s P(s)\phi(s)$, $\langle \phi \rangle_Q = \sum_s Q(s)\phi(s)$. Similarly, we have the variances defined as $\langle (\phi-\langle \phi \rangle)^2\rangle_{P,Q}$, with respect to $P$ and $Q$. Define the KL divergence $D(P|Q)=\sum_s P(s)\ln P(s)/Q(s)$ (defined if $Q(s)=0 \rightarrow P(s)=0$ for any $s$, a property called absolute continuity).

{\bf \emph{Main result -}}  Our main result is the involution Thermodynamic Uncertainty Relation (iTUR) with respect to any $P$ and $Q$, which reads
\begin{equation}
\label{iTUR}
\frac{\langle (\phi - \langle \phi\rangle_P)^2\rangle_P+\langle (\phi - \langle \phi\rangle_Q)^2\rangle_Q}{(1/2)(\langle \phi \rangle_P + \langle \phi \rangle_Q)^2}\geq f\big(\frac{D(P|Q')+D(Q'|P))}{2}\big),
\end{equation}
with $f(x)=\sinh^{-2}(g(x)/2)$ and $g(x)$ is the inverse function of $h(x)=x\tanh(x/2)$, for $x\geq0$, where $P',Q'$ are defined as (\ref{Pprime}) for any involution (\ref{involution}) and any current (\ref{current}) such that $\langle \phi \rangle_P + \langle \phi \rangle_Q \neq 0$.

{\bf \emph{Proof -}} Consider any probabilities $P$ and $Q$, any involution $m$ (\ref{involution}), with corresponding $P'$and $Q'$ (\ref{Pprime}) such that the pairs $P,Q'$ and $Q,P'$ are absolute continuous. Let $\phi$ be any current (\ref{current}). Define $p:S\rightarrow [0,1]$ as
\begin{equation}
\label{probp}
    p(s):=\frac{P(s)+Q'(s)}{2}.
\end{equation}
Note that $\sum_s p(s) = 1$ follows from the normalization of $P$ and $Q'$. Using (\ref{probp}), (\ref{involution}) and (\ref{current}), one gets the average current,
\begin{equation}
\label{property1}
\overline{\phi}:=\langle \phi \rangle_p = \sum_s \phi(s)\frac{(P(s) + Q(s'))}{2} = \frac{\langle \phi \rangle_P - \langle \phi \rangle_Q}{2}.
\end{equation}
Combining (\ref{probp}) and (\ref{property1}), we get a relation for the variance,
\begin{equation}
\label{property2b}
4\langle (\phi-\overline{\phi})^2\rangle_p = 2\langle (\phi - \langle \phi \rangle_P)^2\rangle_P + 2\langle (\phi - \langle \phi \rangle_Q)^2\rangle_Q + (\langle \phi \rangle_P + \langle \phi \rangle_Q)^2.
\end{equation}
Additionally, for every $s$, we have for $P(s)+Q'(s)>0$,
\begin{eqnarray}
\label{property2}
\phi(s)\frac{(P(s) - Q'(s))}{2} =\phi(s)\frac{P(s) - Q'(s)}{P(s)+Q'(s)}p(s),
\end{eqnarray}
and after summing all $s$ on both sides of ((\ref{property2})), using (\ref{involution}) and (\ref{current}), it results in
\begin{eqnarray}
\label{property3}
\frac{\langle \phi\rangle_P + \langle \phi \rangle_Q}{2} = \langle \phi \frac{P - Q'}{P+Q'} \rangle_p = \langle (\phi-\overline{\phi})\frac{P - Q'}{P+Q'} \rangle_p,
\end{eqnarray}
where one should interpret $P,Q'$ inside the averages as $P(s), Q'(s)$ and use that $\langle (P-Q')/(P+Q') \rangle_p=0$. Using Cauchy-Schwarz inequality, one has
\begin{eqnarray}
\label{property4}
\langle (\phi-\overline{\phi})\frac{P - Q'}{P+Q'} \rangle_p^2 \leq \langle (\phi-\overline{\phi})^2 \rangle_p \langle \big(\frac{P - Q'}{P+Q'}\big)^2\rangle_p.
\end{eqnarray}
Now combining (\ref{property3}) and (\ref{property4}), one obtains
\begin{eqnarray}
\label{property5}
\frac{(\langle \phi\rangle_P + \langle \phi \rangle_Q)^2}{4} \leq \langle (\phi-\overline{\phi})^2 \rangle_p \langle \big(\frac{P - Q'}{P+Q'}\big)^2\rangle_p.
\end{eqnarray}
Then, note that 
\begin{equation}
\label{property6}
\frac{P - Q'}{P+Q'}=\frac{P/Q' - 1}{P/Q'+1} = \tanh\big(\frac{1}{2}\ln \frac{P}{Q'}\big).
\end{equation}
Now we use Jensen's inequality, for any $x:S\rightarrow \mathbb{R}$ and any probability function, one has
\begin{equation}
\label{Jensens1}
\langle \tanh(\frac{x}{2})^2 \rangle = \langle \tanh(\frac{g(h(x))}{2})^2 \rangle \leq \tanh(\frac{g(\langle h(x) \rangle)}{2})^2,
\end{equation}
where $h(x):=x\tanh(x/2)$ and $g(x)$ is the inverse function of $h$ for $x\geq0$, $g(h(x))=x$, since $d^2w(h)/dh^2 < 0$, which is a property already used in other contexts \cite{Salazar2021b,Campisi2021,Y.Zhang2019}. Replacing $x=\ln (P/Q')$ and $\langle \rangle = \langle \rangle_p$ in (\ref{Jensens1}) and using (\ref{property6}), it results in
\begin{equation}
\label{Jensens2}
\langle \big(\frac{P - Q'}{P+Q'}\big)^2\rangle_p  \leq \tanh\big(\frac{1}{2}g(\langle h(\ln \frac{P}{Q'}) \rangle_p)\big)^2,
\end{equation}
where the term $\langle h (\ln P/Q')\rangle_p$ can be simplified to
\begin{eqnarray}
\langle h(\ln \big(\frac{P}{Q'}\big)\rangle_p = \sum_s \ln \big(\frac{P(s)}{Q'(s)}) \frac{P(s)-Q'(s)}{P(s)+Q(s)}p(s)
\\
\label{property7}
=\frac{1}{2}(D(P|Q')+D(Q'|P)).
\end{eqnarray}
We finally obtain from (\ref{property5}), 
 (\ref{Jensens2}) and (\ref{property7}),
\begin{equation}
\label{Ineq0}
\frac{(\langle \phi\rangle_P + \langle \phi \rangle_Q)^2}{4} \leq \langle (\phi-\overline{\phi})^2\rangle_p \tanh(\frac{1}{2} g(\frac{{D(P|Q')+D(Q'|P)}}{2}))^2.
\end{equation}
Inverting (\ref{Ineq0}) and using $\tanh(x/2)^{-2}=1+\sin(x/2)^{-2}$, we get
\begin{equation}
\label{Ineq1}
\frac{4\langle (\phi - \overline{\phi})^2\rangle_p}{(\langle \phi \rangle_P + \langle \phi \rangle_Q)^2}\geq 1+ f\big(\frac{D(P|Q')+D(Q'|P))}{2}\big).
\end{equation}
Using (\ref{property2b}) in (\ref{Ineq1}), we obtain our main result (\ref{iTUR}).

{\bf \emph{Discussion -}} In the derivation of (\ref{iTUR}), we did not use any FT. Instead, the involution (\ref{involution}) was the only assumption needed. If one thinks of $s$ as a process, the assumption is simply stating the existence of a single inverse process $m(s)$, such that $m(m(s))=s$, which is the simplest property behind the arrow of time: if you flip it twice, you get the original direction.

In the derivation above, we used several ideas from recent literature on TURs. Specifically, the analogous of definition (\ref{property1}) appeared in the asymmetric TUR \cite{Proesmans2019,Potts2019,Gianluca2022,Nishiyama2020,VanTuan2020}, and the tightest Jensen's inequality in (\ref{Jensens1}) was also used before \cite{Timpanaro2019b,Y.Zhang2019,Salazar2022b}. Our contribution was to write a general TUR solely in terms of the involution $m$ instead of any FT, which is a mathematical statement about the object $(S,P,Q)$ for any $m$ and $\phi$. The relation between iTUR and the physical TURs (in terms of entropy productions) will be explored below. We also noted a general expression similar to (\ref{iTUR}) appeared recently \cite{Nishiyama2022}, where the equivalence can be obtained considering in our notation $S=\mathbb{R}$, and for any $s=x \in \mathbb{R}$, $\phi(x)=x$, $m(x)=-x$.

{\bf \emph{Application 1: exchange FT -}}
Consider $p(\sigma,\phi)$ as the probability of observing the entropy production $\sigma$ an the current $\phi$. In this case, the exchange fluctuation theorem reads
\begin{equation}
\label{xFT}
p(\sigma,\phi)=p(-\sigma,-\phi)e^{\sigma},
\end{equation}
valid for a series of setups \cite{Jarzynski2004a,Timpanaro2019b,Seifert2005}, also called the strong detailed fluctuation theorem \cite{Merhav2010,Seifert2012Review}, the Evan-Searles fluctuation theorem \cite{Evans2002} and Gallavotti-Cohen relation \cite{Gallavotti1995}. To obtain the corresponding TUR from the main result (\ref{iTUR}), we use $S=\{(q,\sigma)|q, \sigma \in \mathbb{R}\}$, also $Q=P$ and $m(\phi,\sigma)=(-\phi,-\sigma)$ as the involution, which makes $D(P|Q')=D(Q'|P)=D(P|P')=\langle \sigma \rangle$ from (\ref{xFT}). In this case, (\ref{iTUR}) results in the following exchange thermodynamic uncertainty relation,
\begin{equation}
\label{xTUR}
\frac{\langle \phi - \overline{\phi}\rangle^2}{\langle \phi \rangle^2} \geq f(\langle \sigma \rangle),
\end{equation}
famously derived from the exchange FT (\ref{xFT}) \cite{Merhav2010,Hasegawa2019,Timpanaro2019b}. Again, note that (\ref{xTUR}) followed immediately from (\ref{iTUR}), where the role of the FT was only to assign $D(P|P')=\langle \sigma \rangle$.

{\bf \emph{Application 2: asymmetric FT -}} The detailed fluctuation theorem reads
\begin{equation}
\label{DFT}
P_F(\Gamma)=P_B(\Gamma^\dagger)e^{\sigma(\Gamma)},
\end{equation}
which is the definition of stochastic entropy production \cite{Seifert2005} for a trajectory $\Gamma=\{(x_0,\lambda_0),(x_1,\lambda_1),...,(x_N,\lambda_N)\}$ in terms of forward (F) and backward (B) probabilities, $P_F$ and $P_B$, where $\Gamma^\dagger=\{(x_N,\lambda_N),...,(x_0,\lambda_0)\}$ is the inverse trajectory, $x_i$ is the state of the system at time $t_i$ and $\lambda_i$ is a controllable parameter at time $t_i$.
The application of (\ref{iTUR}) is straightforward by considering $s=\Gamma$, the involution $m(\Gamma)=\Gamma^\dagger$, and probabilities $P(s)=P_F(\Gamma)$, $Q(s)=P_B(\Gamma)$. In this case, $Q'(s)=Q(m(s))=P_B(\Gamma^\dagger)$. We obtain from (\ref{iTUR}) for any current $\phi(\Gamma)=-\phi(\Gamma^\dagger)$:
\begin{equation}
\label{aTUR}
\frac{\langle (\phi - \langle \phi\rangle_F)^2\rangle_F+\langle (\phi - \langle \phi\rangle_B)^2\rangle_B}{(1/2)(\langle \phi \rangle_F + \langle \phi \rangle_B)^2}\geq f\big(\frac{\overline{\sigma} + \overline{\sigma}_B}{2}\big),
\end{equation}
where $\sigma(\Gamma):=\ln(P_F(\Gamma)/P_B(\Gamma)$, $\sigma_B(\Gamma):=-\sigma(\Gamma^\dagger)$, $\overline{\sigma}:=\langle \sigma(\Gamma) \rangle_F = \sum_\Gamma P_F(\Gamma)\ln[P_F(\Gamma)/P_B(\Gamma^\dagger)]$,  $\overline{\sigma}_B:=\langle -\sigma(\Gamma^\dagger) \rangle_B = \sum_\Gamma P_B(\Gamma) \ln [P_B(\Gamma)/P_F(\Gamma^\dagger)]$ and the subscripts $F,B$ are the averages over $P_F(\Gamma),P_B(\Gamma)$. As in the case of the exchange TUR, the role of the FT was to assign $\overline{\sigma}=D(P_F|P_B')$ and $\overline{\sigma}_B=D(P_B'|P_F)=D(P_B|P_F')$, where $P_F'(\Gamma):=P_F(\Gamma^\dagger),P_B'(\Gamma):=P_B(\Gamma^\dagger)$. Asymmetric TURs were the subject of recent results \cite{Proesmans2019,Potts2019,Gianluca2022}, notably because they are applicable to asymmetric protocols. Note that the symmetric case $F=B$ recovers a form similar to the exchange TUR (\ref{xTUR}). The case $\langle \phi \rangle_F=\langle \phi \rangle_B \neq 0$ reproduces the asymmetric TUR of \cite{Gianluca2022}.

Equivalently, for completeness, note that the asymmetric problem can be mapped in the same domain of the exchange TUR, $S=\{(q,\sigma)|q, \sigma \in \mathbb{R}\}$, by considering
\begin{eqnarray}
\label{fprob}
p(\sigma,\phi):=\sum_\Gamma P_F(\Gamma)\theta(\phi(\Gamma)-\phi)\theta(\sigma(\Gamma)-\sigma),
\\
\label{fprob2}
\overline{p}(\sigma,\phi):=\sum_\Gamma P_B(\Gamma)\theta(\phi(\Gamma)-\phi)\theta(\sigma_B(\Gamma)-\sigma),
\end{eqnarray}
with $\theta(0)=1$ and $\theta(x)=0$ if $x\neq 0$, which makes the DFT (\ref{DFT}) valid, $p(\sigma,\phi)=\overline{p}(-\sigma,-\phi)e^\sigma$, and it recovers (\ref{aTUR}) in 
terms of $p$ (\ref{fprob}) and $\overline{p}$ (\ref{fprob2}), replacing $\langle \rangle_F = \langle \rangle_p$, $\langle \rangle_B=\langle \rangle_{\overline{p}}$, $\overline{\sigma}=\sum_\sigma \sigma p(\sigma, \phi)$ and $\overline{\sigma}_B=\sum_\sigma \sigma \overline{p}(\sigma,\phi)$.

{\bf \emph{Application 3: Total Variation bound -}} In this application, we consider any set $S$, probabilities $Q=P$ and involution $m$ for the particular current
\begin{equation}
\label{sign}
\phi(s):=\sign[P(s)-P(s')].
\end{equation}
From (\ref{sign}), we have $\phi(s')=\sign[P(s')-P(s)]=-\phi(s)$, as expected. The mean of $\phi$ is given by
\begin{equation}
\langle \phi \rangle_P = \sum_s \sign[P(s)-P(s')]P(s)= \frac{1}{2}\sum_s |P(s)-P(s')|, 
\end{equation}
where we used $\sum_s \phi(s)P(s)=(1/2)\sum_{s}\phi(s)(P(s)-P(s'))$, since $s'=m(s)$ is an involution. Note that $\Delta(P,P')=(1/2)\sum_s|P(s)-P(s')|$ is the definition of the total variation distance between $P$ and $P'$. The variance of $\phi$ is
\begin{equation}
\label{varTV}
\langle \phi^2 \rangle_P - \langle \phi \rangle_P^2 = 1-P_0 - \Delta(P,P')^2,
\end{equation}
where defined the parameter
\begin{equation}
\label{eps}
P_0 := \sum_{s} P(s)\theta(P(s)-P(s')),
\end{equation}
which one might understand intuitively as the probability of drawing a process from equilibrium ($P(s)=P(s')$, or detailed balance). Using (\ref{sign}) and (\ref{eps}) in (\ref{iTUR}) for $P=Q$, one obtains
\begin{equation}
\label{TVbound}
\frac{1-P_0 - \Delta(P,P')^2}{\Delta(P,P')^2}\geq f\big(D(P|P')\big),
\end{equation}
which inverts to
\begin{equation}
\label{TVbound2}
D(P|P') \geq 2\frac{\Delta(P,P')}{\sqrt{1-P_0}}\tanh^{-1}\big(\frac{\Delta(P,P')}{\sqrt{1-P_0}}\big),
\end{equation}
which is a lower bound for the Kullback-Leibler divergence for any $P$, when $P$ and $P'$ are connected by any involution, $P'(s)=P(m(s))$. We also note that $D(P|P')\geq B(\Delta(P,P')/\sqrt{1-P_0}) \geq B(\Delta(P,P'))$, for $B(x):=2x\tanh^{-1}(x)$, since $dB(x)/dx>0$, for $x>0$ and from (\ref{varTV}) $1 \geq \Delta(P,P')/\sqrt{1-P_0}\geq \Delta(P,P')$, which improves on the lower bound of \cite{Salazar2022b} when $P_0$ is known. A similar relation was also used to obtain a lower bound for the apparent violations of the second law \cite{Salazar2021b} in terms of the entropy production, when $D(P|P')=\langle \sigma \rangle$.

{\bf \emph{Application 4: Moment generating function -}} 
We apply the iTUR to obtain a general relation for the moment generating function (mgf) of the random variable $\ln(P(s)/P(s'))$. First, consider again $P=Q$ and any involution $m$. Define the following moment generating function for the variable $\ln(P(s)/P'(s))$ given by $G(\alpha)=\langle \exp(\alpha\ln(P/P'))\rangle_P=\langle (P/P')^\alpha\rangle_P$. The first and second derivatives of $G(\alpha)$ yield $G_0':=G'(0)=\langle \ln P/P' \rangle_P = D(P|P').$ and $G''(0)=\langle (\ln P/P')^2\rangle$. The iTUR for the current $\phi(s)=\ln P(s)/P'(s)$ yields the bound
\begin{equation}
\label{mgf0}
\frac{G_0'^2}{G''(0)} \leq \tanh(\frac{1}{2}g(G'_0))^2
\end{equation}
However, one gets a more general relation for the mgf $G(\alpha)$ considering the current
\begin{equation}
\label{mgf1}
\phi(s):=(\frac{P(s)}{P'(s)})^\alpha-(\frac{P'(s)}{P(s)})^\alpha,
\end{equation}
for $P(s),P(s')\neq 0$ and $\phi(s)=0$ otherwise. Check that $\phi(s')=-\phi(s)$ as expected. The first and second moments of $\phi$ read
\begin{eqnarray}
\label{mgf2}
\langle \phi \rangle_P = \sum_s\phi(s)P(s)=G(\alpha)-G(-\alpha),
\\
\label{mgf3}
\langle \phi^2 \rangle_P = \sum_s\phi(s)^2P(s)=G(2\alpha)+G(-2\alpha)-2.
\end{eqnarray}
Inserting (\ref{mgf2}) and (\ref{mgf3}) in (\ref{iTUR}), we obtain
\begin{equation}
\label{mgf4}
\frac{[G(\alpha)-G(-\alpha)]^2}{[G(2\alpha)+G(-2\alpha)-2]}\leq \tanh^2\big(\frac{1}{2}g(G'_0)\big).
\end{equation}
In the specific case of $\alpha\rightarrow 0$, (\ref{mgf4}) yields (\ref{mgf0}). However, (\ref{mgf4}) is more general, as it holds for finite values of $\alpha$. Also note that $(1-\alpha)D_\alpha(P|P')=\ln(G(\alpha-1))$ defines the Rényi divergence $D_\alpha(P|P')$.

{\bf \emph{Application 5: Fluctuation-response bound -}} Let $p(x|\theta)$ be a probability density function in $\mathbb{R}$ for some parameter $\theta$. We define $S=\mathbb{R}$, $m(x)=-x$, $P(x)=p(x|\theta+\epsilon)$ and $Q(x)=p(-x|\theta)$, so that any asymmetric current, $\phi(-x)=-\phi(x)$, results in the averages $\langle \phi \rangle_{P}=\int \phi(x)p(x|\theta+\epsilon)dx:=\langle \phi \rangle_{\theta+\epsilon}$ and $\langle \phi \rangle_{Q}=\int \phi(x)p(-x|\theta)dx=-\langle \phi \rangle_{\theta}\rangle$. In this case the iTUR (\ref{iTUR}) reads

\begin{equation}
\label{CramerRao}
\frac{(1/2)(\langle \phi \rangle_{\theta+\epsilon} - \langle \phi \rangle_{\theta})^2}{\langle (\phi - \langle \phi\rangle_{\theta+\epsilon})^2\rangle_{\theta+\epsilon}+\langle (\phi - \langle \phi\rangle_{\theta})^2\rangle_{\theta}}\leq \sinh\big(\frac{g(\Sigma(p_{\theta+\epsilon},p_{\theta}))}{2}\big)^2,
\end{equation}
where we defined $\Sigma(p_{\theta+\epsilon},p_{\theta}):=(1/2)\int [p(x|\theta+\epsilon)-p(x|\theta)]\ln (p(x|\theta+\epsilon)/p(x|\theta))dx \geq 0$.

Note that (\ref{CramerRao}) is a type of fluctuation-response inequality \cite{Dechant2020}. In the limiting case $\epsilon \rightarrow 0$, it reduces to the Crámer-Rao inequality as follows: expanding (\ref{CramerRao}) in $\epsilon$, one obtains $\langle \phi \rangle_{\theta+\epsilon} - \langle \phi \rangle_\theta = \partial_\theta \langle \phi \rangle_\theta \epsilon + \mathcal{O}(\epsilon^2)$ and $\langle (\phi - \langle \phi \rangle_{\theta+\epsilon})^2\rangle_{\theta+\epsilon} = \langle (\phi - \langle \phi \rangle_{\theta})^2\rangle_{\theta} + \mathcal{O}(\epsilon)$. We also have $\Sigma(p_{\theta+\epsilon}, p_{\theta})=(\epsilon^2/2)\int (\partial_\theta p(x|\theta))^2/p(x|\theta) dx + \mathcal{O}(\epsilon^3)= (\epsilon^2/2)I(\theta)+\mathcal{O}(\epsilon^3)$, where $I(\theta)$ is the Fisher information. Using $g(x)\approx \sqrt{2x}$ and $\sinh(x)\approx x$ for $x\approx 0$, one has $\sinh(g(\Sigma)/2)^2\approx \Sigma/2 \approx \epsilon^2 I(\theta)/4$, and (\ref{CramerRao}) results in
\begin{equation}
\label{CramerRao2}
\frac{(\partial_{\theta}\langle \phi \rangle_{\theta} )^2}{\langle (\phi - \langle \phi\rangle_{\theta})^2\rangle_{\theta}}\leq I(\theta),
\end{equation}
in the limit $\epsilon\rightarrow 0$, which is the famous Crámer-Rao bound \cite{Hasegawa2019b,Ito2020}. 

{\bf \emph{Conclusions -}}
We showed that a set $S$ equipped with an involution $m$ is able to produce the involution Thermodynamic Uncertainty Relation (iTUR) for any pair of probabilities $P,Q$ and any asymmetric current $\phi(m(s))=-\phi(s)$. Remarkably, the FT was not used in the derivation, which might sound unusual at first glance when compared to other TURs. With that result, we argue that the origin of the TUR is better understood as a consequence of the involution, where each event $s$ has a single inverse $s'$. In other words, the arrow of time is a two-way street: reversing the direction twice takes you back to the initial direction, $m(m(s))=s$. This apparently naive property holds the key to the iTUR and the underlying applications. This is a purely nonequilibrium result, as a system in equilibrium ($P(s)=Q'(s)$, equivalent to detailed balance) would collapse the relations obtained in this paper.

As applications, we showed how the result implies the exchange fluctuation theorem and a tight form of the asymmetric fluctuation theorem, both important results in nonequilibrium thermodynamics. We also showed how the iTUR is related to a connection between $D(P|P')$ and other statistics of $P$ and $P'$, such as the total variation distance $\Delta(P,P')$, the mgf $\langle \exp(\alpha \ln(P/P')\rangle$ and a fluctuation-response inequality, combining different results from nonequilibrium thermodynamics under the same framework.

\bibliography{lib6}
\end{document}